\documentclass[twocolumn,aps,prb,showpacs]{revtex4-1}
\usepackage{amsfonts}
\usepackage{amsmath}
\usepackage{graphicx}
\usepackage{bm}
\usepackage{amssymb}
\usepackage{dcolumn}
\usepackage{color}

\begin{document}

\title{Enhanced Magneto-optical Kerr Effect at Fe/Insulator Interfaces}

\author{Bo Gu$^{1}$, Saburo Takahashi$^{2}$, and Sadamichi Maekawa$^{1,3}$} 
 
\affiliation{$^1$ Advanced Science Research Center, Japan Atomic Energy Agency, Tokai 319-1195, Japan \\   
$^2$ Institute for Materials Research, Tohoku University, Sendai 980-8577, Japan \\      
$^3$ ERATO, Japan Science and Technology Agency, Sendai 980-8577, Japan}                            

\date{\today}
\begin{abstract}
Using density functional theory calculations, we have found an enhanced magneto-optical Kerr effect in Fe/insulator interfaces. The results of our study indicate that interfacial Fe atoms in the Fe films have a low-dimensional nature, which causes the following two effects: (i) The diagonal component $\sigma_{xx}$ of the optical conductivity decreases dramatically because the hopping integral for electrons between Fe atoms is suppressed by the low dimensionality. (ii) The off-diagonal component $\sigma_{xy}$ of the optical conductivity does not change at low photon energies, but it is enhanced at photon energies around 2 eV, where we obtain enhanced orbital magnetic moments and spin-orbit correlations for the interfacial Fe atoms. A large Kerr angle develops in proportion to the ratio $\sigma_{xy}/\sigma_{xx}$. Our findings indicate an efficient way to enhance the effect of spin-orbit coupling at metal/insulator interfaces without using heavy elements.     
\end{abstract}

 \maketitle
\section{Introduction}
Magneto-optical Kerr effect (MOKE) is the phenomenon in which the light reflected from a magnetized material has a rotated plane of polarization.
Given its sensitivity, local probing nature, and experimental simplicity, this phenomenon has significantly impacted research on magnetic
materials~\cite{MOKE-Bader}. The MOKE originates from spin-orbit coupling in materials and has been extensively studied.
It has also been applied for magneto-optical data recording~\cite{MOKE-Data}. Recently, the effect has played a significant role in the rapidly developing field of spintronics. For example, the first experimental observation of the spin Hall effect~\cite{MOKE-SHE}, direct experimental observation of the skyrmion Hall effect~\cite{MOKE-SKX-Hall}, and investigations of spin-orbit torques in metallic and insulating magnetic bilayers~\cite{MOKE-SOT-1, MOKE-SOT-2, MOKE-SOT-3} were all performed using the MOKE. The effect is not limited to ferromagnetic materials, and a substantial MOKE has been proposed recently for some antiferromagnetic materials from theoretical calculations~\cite{MOKE-AF-1, MOKE-AF-2}.  

For applications to magneto-optical devices, a substantial MOKE is required; this depends on the availability of materials with considerable spin-orbit coupling. Because it is a relativistic effect, spin-orbit coupling is small in many materials, but it is substantial in heavy elements such as Pt. To obtain large spin-orbit coupling, many different effects have recently been applied. These effects are as follows: spin-orbit splitting in the band structure due to the Rashba effect in systems without inversion symmetry~\cite{LSOC-Rashba-Fert, LSOC-Rashba-Zhang,LSOC-Rashba-Barnes}; charge-spin conversions due to spin-momentum locking of topological surface states~\cite{LSOC-TI-Lock-Fan, LSOC-TI-Lock-Shiomi, LSOC-TI-Lock-Ralph, LSOC-TI-Lock-Otani}; Dzyaloshinskii--Moriya interactions in bilayer systems with heavy elements~\cite{LSOC-Heavy-Fert,LSOC-Heavy-Bode, LSOC-Heavy-Heinze, LSOC-Heavy-Emori, LSOC-Heavy-Luchaire}; spin-orbit couplings due to impurities~\cite{LSOC-Imp-Guo, LSOC-Imp-AuFe, LSOC-Imp-AuPt, LSOC-Imp-CuBi-Niimi, LSOC-Imp-CuBi-Gu, LSOC-Imp-CuIr}.       

There are a few ways to enhance the MOKE. One is to employ alloys~\cite{Kerr-Sugimoto,DFT-Kerr-Osterloh} or multilayers~\cite{DFT-Kerr-Guo,DFT-Kerr-Ebert} of transition metals such as Fe or Co and heavy elements such as Pt because heavy elements have large spin-orbit coupling. 
Another way to enhance the MOKE is to utilize the plasma edge effect~\cite{DFT-Kerr-Feil-Haas}. By decreasing the diagonal component $\sigma_{xx}$ of the optical conductivity, the MOKE is enhanced because the Kerr rotation angle is proportional to $1/\sigma_{xx}$. Such a mechanism has been discussed to enhance the MOKE in Fe/Cu bilayers~\cite{Kerr-Katayama}. The photonic effect can also be used to enhance the MOKE because of multiple interference of light within the magnetic multilayers~\cite{Kerr-photonic-Inoue, Kerr-photonic-Fedyanin, Kerr-photonic-Koba}. Cavity enhancement of the MOKE has also been reported, in which a dielectric layer acts as a Fabry--Perot optical cavity~\cite{Kerr-cavity-2004, Kerr-cavity-2014, Kerr-cavity-2017}. Metamaterials, i.e., composite nanostructured materials, can also show enhancement of the MOKE~\cite{Kerr-meta-2007, Kerr-meta-2014, Kerr-meta-2017}.

\begin{figure}[tbp]
\includegraphics[width = 8.5 cm]{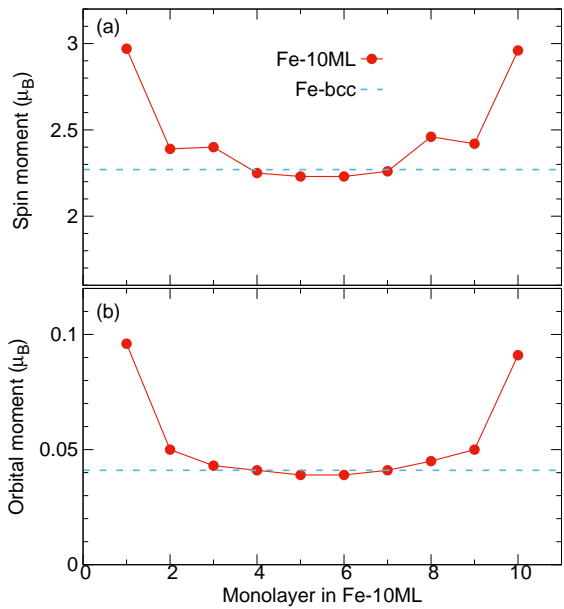}
\caption{(a) The spin moment and (b) the orbital moment of a ten-monolayer supercell of Fe (Fe-10ML) as obtained by density functional theory calculations. 
The 1st-ML and the 10th-ML are Fe/vacuum interfaces. We performed structural relaxation in the calculation.   
For reference, the calculated value for bulk body-centered cubic iron (Fe-bcc) is shown by the dashed line. }
\label{F-spin-orbit-fe}
\end{figure}

In this paper, we propose an efficient way to enhance the MOKE using Fe/insulator interfaces. 
In contrast to previous methods, this new method does not use heavy elements. Instead, it focuses on spin-orbit interactions, which can be enhanced by our method, while they have been ignored in previous applications of plasmonic or photonic effects.
We were motivated by the discovery of large magnetic anisotropies in ultrathin Fe/MgO films~\cite{Fe-MgO-Li, Fe-MgO-Boubeta,Fe-MgO-Beltran, Fe-MgO-Baumann, Fe-MgO-Bose}. The Fe/MgO interface is used in a wide range of devices, and enhanced spin-orbit interactions are expected because of the large magnetic anisotropy. In this paper, we use density functional theory calculations to show that atomically thin Fe layers as well as Fe/insulator multilayers comprising a few Fe layers and a few insulating layers of MgO or AlF$_3$ can produce a large MOKE and large spin-orbit interactions. Our results indicate an efficient way to obtain enhanced spin-orbit coupling effects at metal/insulator interfaces without using heavy elements.             

\section{Density functional theory calculation results}

We first discuss the spin and orbital magnetic moments at a metallic Fe surface.   
Figure \ref{F-spin-orbit-fe} shows the monolayer (ML)-resolved spin and orbital moments of a ten-monolayer Fe supercell (Fe-10ML) at the Fe (001) surface. 
The 1st-ML and the 10th-ML are the two surface layers (Fe/vacuum interfaces), and we take the vacuum along the [001] direction to be 6ML thick. We performed density functional theory calculations using the \texttt{WIEN2K} package~\cite{DFT-Wien2k}. 
In the calculations, the exchange-correlation interactions are described by the Perdew--Burke--Ernzerhof generalized gradient approximation~\cite {DFT-PBE}, and spin-orbit coupling is included using the second-variation method~\cite{DFT-SOC}. We used the atomic sphere radius parameter RMT = 2.24 for Fe, and employed the cutoff parameter RK$_{max}$ = 9. We used 12 $\times$ 12 $\times$ 1 k-point sampling for the Fe-10ML calculation. The magnetization direction is along the [001] direction. We performed structural relaxation for bulk body-centered cubit iron (Fe-bcc) and for Fe-10ML. In the Fe-10ML supercell, the lattice of MLs Nos. 1--4 was fixed to be equal to that of bulk Fe-bcc, the z component of the lattice of MLs Nos. 5--10 was relaxed, and the in-plane components of the lattice were fixed as those of bulk Fe-bcc.
We found the spin and orbital magnetic moments to be strongly enhanced in comparison to the values of bulk Fe-bcc. While this enhancement is limited mainly to the two surface layers, the spin and orbital moments of the internal layers decrease dramatically, thus approaching the value of bulk Fe, as shown in Fig. \ref{F-spin-orbit-fe}. This suggests that the surface Fe atoms are in a low-dimensional state intermediate between that of bulk Fe, which has small spin and orbital moments, and that of a single Fe atom state, for which the 3$d^6$ state has rather large spin and orbital moments. 
Our result is in good agreement with previous calculations pertaining to the spin and orbital moments of 3$d$ transition metals in both surface and bulk cases~\cite{Bruno, Eriksson}.

Given that the MOKE originates from spin-orbit coupling, we investigated whether the enhanced spin and orbital moments at the Fe surface can induce a substantial Kerr effect. The Kerr angle is given as follows:
\begin{equation}
  \theta_{Kerr}\left(\omega\right) = -{\text Re}\frac{\epsilon_{xy}}{(\epsilon_{xx}-1)\sqrt{\epsilon_{xx}}},
   \label{E-kerr}
\end{equation}
where $\epsilon_{xx}$ and $\epsilon_{xy}$ are the diagonal and off-diagonal components of the dielectric tensor $\epsilon$, and $\omega$ is the photon energy, respectively. The dielectric tensor $\epsilon$ and the optical conductivity tensor $\sigma$ are related as follows: 
\begin{equation}
\sigma\left(\omega\right) =\frac{\omega}{4\pi i}\left[\epsilon\left(\omega\right)-I\right],
\label{E-sigma} 
\end{equation}
where $I$ is the unit tensor. 
We performed density functional theory calculations with the \texttt{QUANTUM ESPRESSO} package~\cite{DFT-QE} along with maximally localized Wannier function calculations using the \texttt{wannier90} tool~\cite{DFT-W90} to obtain the optical conductivity tensor $\sigma$ and the Kerr angle $\theta_{Kerr}$ for the following three different cases: bulk Fe-bcc, Fe-10ML, and Fe-2ML. 
In the calculations, the exchange-correlation interactions are described by the Perdew--Burke--Ernzerhof generalized gradient approximation, electron-ion interactions are represented by the Rabe--Rappe--Kaxiras--Joannopoulos ultrasoft pseudopotential, and spin-orbit coupling is included~\cite{QE-UPF}. The kinetic energy cutoff parameters for the wavefunctions (ecutwfc) and for the charge density and potential (ecutrho) are taken to be ecutwfc = 60 Ry and ecutrho = 600 Ry for Fe-bcc, ecutwfc = 60 Ry and ecutrho = 800 Ry for Fe-10ML, and ecutwfc = 60 Ry and ecutrho = 1200 Ry for Fe-2ML. In the self-consistent calculations, the k-point samplings were 16 $\times$ 16 $\times$ 16 for Fe-bcc, 12 $\times$ 12 $\times$ 1 for Fe-10ML, and 12 $\times$ 12 $\times$ 6 for Fe-2ML. To evaluate the optical conductivity tensor in the Wannier functional calculations, we used 25 $\times$ 25 $\times$ 25 k-point sampling. The magnetization direction is along the [001] direction. We used the structures previously obtained for bulk Fe-bcc and Fe-10ML. For the Fe-2ML calculation, we took the vacuum along the [001] direction to be 2ML thick, relaxed the z component of the lattice, and fixed the in-plane components of the lattice to be equal to those of bulk Fe-bcc.

\begin{figure}[tbp]
\includegraphics[width = 8.5 cm]{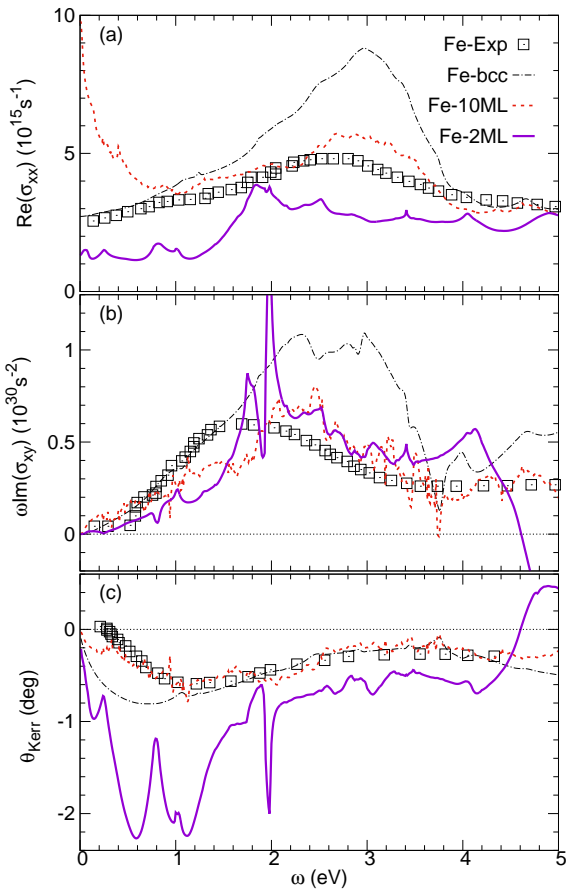}
\caption{(a) The real part of the diagonal component Re($\sigma_{xx}$) of the optical conductivity; (b) the imaginary part of the off-diagonal component Im($\sigma_{xy}$) of the optical conductivity multiplied by the photon energy $\omega$; (c) the Kerr angle $\theta_{Kerr}$, each as a function of the photon energy $\omega$. Density functional theory calculations for the following three cases are shown: bulk Fe-bcc, Fe-10ML, and Fe-2ML.
The experimental values of Re($\sigma_{xx}$) \cite{Exp-Fe-xx}, $\omega$Im($\sigma_{xy}$) \cite{Exp-Fe-xy}, and $\theta_{Kerr}$ \cite{Exp-Fe-Kerr} for bulk Fe-bcc are shown for comparison.}
\label{F-kerr-surf}
\end{figure}

As shown in Figs. \ref{F-kerr-surf}(a)--(c), for the bulk Fe-bcc case, our results are in good agreement with experimental measurements of the diagonal component $\sigma_{xx}$ of the optical conductivity~\cite{Exp-Fe-xx}, the off-diagonal component $\sigma_{xy}$ of the optical conductivity~\cite{Exp-Fe-xy}, and the Kerr angle $\theta_{Kerr}$~\cite{Exp-Fe-Kerr}. Moreover, our results are consistent with the previous calculations of $\sigma$ and $\theta_{Kerr}$ for the case of bulk Fe-bcc~\cite {DFT-Kerr-Oppeneer, DFT-Kerr-Guo, DFT-Kerr-Ebert, DFT-Kerr-Mainkar, DFT-Kerr-Delin, DFT-Kerr-Yao, DFT-Kerr-Yates, DFT-Kerr-Sangalli}. Interestingly, compared to the bulk Fe-bcc case the calculated results for $\sigma_{xx}$, $\sigma_{xy}$, and $\theta_{Kerr}$ for the Fe-10ML case are closer to the experimental values, as shown in Figs. \ref{F-kerr-surf}(a)--(c). We note that intraband transitions are not included in our calculations. The intraband contribution, as represented by the Drude formula, has usually been added to obtain good comparisons between the numerical and experimental results for the MOKE in metallic Fe, Co, and Ni using the Drude parameters extracted from experimental data~\cite{DFT-Kerr-Delin}. Interestingly, the good agreement between our calculations for the Fe-10ML case and the experiments provides an alternative way to understand the Kerr experiment in metallic Fe when the Drude contribution is not included. In addition, the Fe-10ML case is still dominated by Fe atoms with bulk-like properties, as shown in Fig. \ref{F-spin-orbit-fe}; thus there is no enhancement of $\theta_{Kerr}$ in the Fe-10ML case.

These findings for the Fe-10ML case led us to consider the Fe-2ML case, which consists entirely of surface Fe atoms, i.e., the two monolayers are the two surfaces (Fe/vacuum interfaces). As shown in Fig. \ref{F-kerr-surf}(c), quite a large Kerr angle $\theta_{Kerr}$ is obtained for the Fe-2ML case, particularly for photon energies $\omega$ below 2 eV. Two factors rooted in the Fe surface produce this large Kerr angle. 
First, the diagonal component $\sigma_{xx}$ of the optical conductivity decreases dramatically, as shown in Fig. \ref{F-kerr-surf}(a), 
because the hopping integral for electrons between Fe atoms is suppressed owing to the low dimensionality of the atomically thin Fe layers. 
Second, the off-diagonal component $\sigma_{xy}$ of the optical conductivity, in contrast to $\sigma_{xx}$, does not change much for photon energies $\omega<$  1eV, and it even increases above the value of $\sigma_{xy}$ for bulk Fe at photon energies close to $\omega \sim$ 2eV, as shown in Fig. \ref{F-kerr-surf}(b). 
According to Eqs. (\ref{E-kerr}) and (\ref{E-sigma}), the large values of $\sigma_{xy}/\sigma_{xx}$ and $\theta_{Kerr}$ are obtained because of decreased $\sigma_{xx}$ and large $\sigma_{xy}$. 

\begin{figure}[tbp]
\includegraphics[width = 8.5 cm]{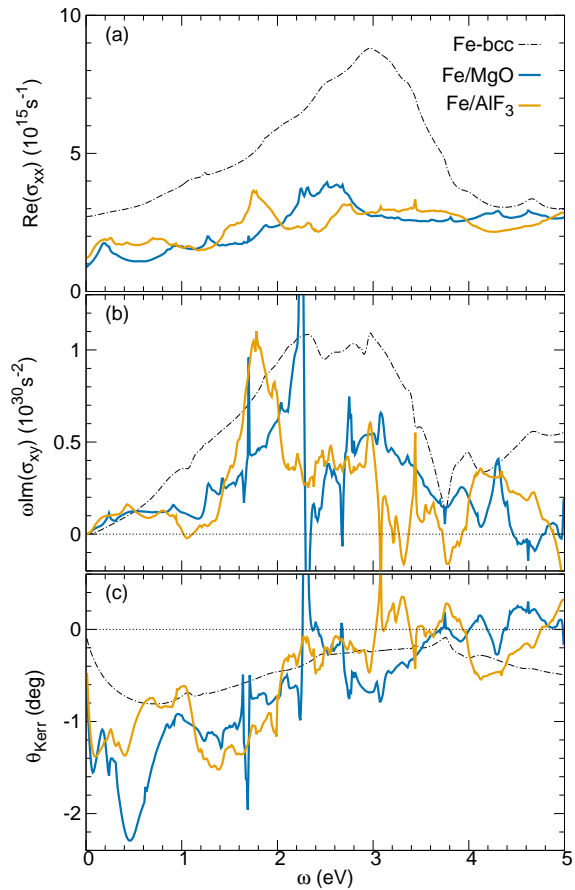}
\caption{(a) The real part of the diagonal component Re($\sigma_{xx}$) of the optical 
conductivity; (b) the imaginary part of the off-diagonal 
component Im($\sigma_{xy}$) of the optical conductivity multiplied by the photon energy $\omega$; 
(c) the Kerr angle $\theta_{Kerr}$, each as a function of 
the photon energy $\omega$, for Fe/MgO and Fe/AlF$_3$ multilayers, where the 
Fe layer is two monolayers thick. The calculated results for bulk Fe-bcc are shown for reference.}
\label{F-kerr-2ml}
\end{figure}
   
Motivated by the enhanced Kerr effect we found for the Fe-2ML case, we continued with studies of other Fe/insulator interfaces. To maintain the large Kerr effect we obtained for an Fe layer comprising solely interfacial Fe atoms, we fixed the Fe layer to be two monolayers thick. 
For the Fe/insulator multilayers, we added 2ML of MgO or AlF$_3$ along the [001] direction as the insulator layers. We again relaxed the z component of the lattice, and we fixed the in-plane components of the lattice to be equal to those of bulk Fe-bcc. As shown in Fig. \ref{F-kerr-2ml}(c), we obtained a large Kerr angle, particularly for photon energies $\omega$ below 2 eV owing to the same mechanism as in the previous Fe-2ML case, i.e., the diagonal component $\sigma_{xx}$ of the optical conductivity decreases dramatically, as shown in Fig. \ref{F-kerr-2ml}(a), while the off-diagonal component $\sigma_{xy}$ of the optical conductivity does not change for photon energies lower than 0.5 eV, and it even increases above the value for bulk Fe at photon energies around 2 eV, as shown in Fig. \ref{F-kerr-2ml}(b). These calculations indicate that the enhancement of the Kerr effect in Fe/insulator multilayers is rather robust against the choice of insulator.               

\begin{figure}[tbp]
\includegraphics[width = 8.5 cm]{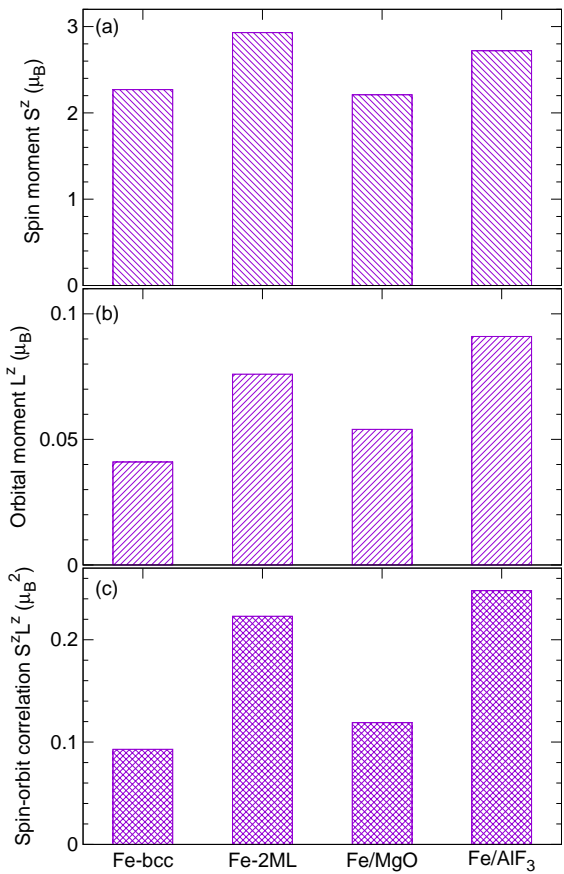}
\caption{(a) Spin moment $S^z$, (b) orbital moment $L^z$, and (c) spin-orbit correlation $S^zL^z$ for Fe-2ML, Fe/MgO and Fe/AlF$_3$ multilayers 
calculated using density functional theory. The structures are the same as those in Figs. \ref{F-kerr-surf} and \ref{F-kerr-2ml}. The calculated results for bulk Fe-bcc are shown for reference. }
\label{F-spin-orbit-2ml}
\end{figure}

Since $\sigma_{xy}$ originates from spin-orbit coupling in the systems, we studied the spin and orbital magnetic moments and their correlations for the Fe-2ML, Fe/MgO, and Fe/AlF$_3$ multilayers, for which the structures are the same as those mentioned above. 
As shown in Fig. \ref{F-spin-orbit-2ml}(a), the spin moments $S^z$ of Fe-2ML and of Fe/AlF$_3$ increase to $\sim$ 3 $\mu_B$, while those of Fe/MgO remain close to 2.2 $\mu_B$, which is the calculated value for bulk Fe. As shown in Fig. \ref{F-spin-orbit-2ml}(b), the orbital moment $L^z$ increases in all three cases. Consequently, we define an effective spin-orbit correlation, as shown in Fig. \ref{F-spin-orbit-2ml}(c), which is the product of $S^z$ in Fig. \ref{F-spin-orbit-2ml}(a) and $L^z$ in Fig. \ref{F-spin-orbit-2ml}(b). The spin-orbit correlation $S^zL^z$ increases for all three cases.    
 	
\begin{figure}[tbp]
\includegraphics[width = 8.5 cm]{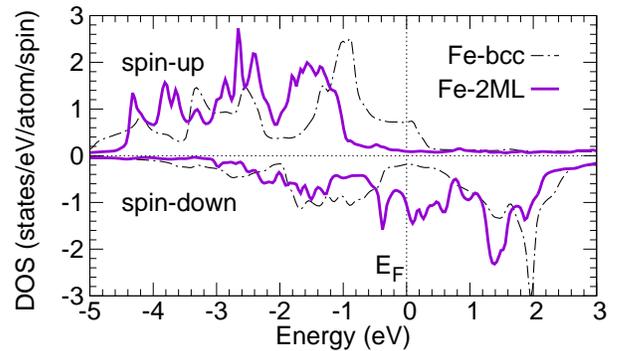}
\caption{ Density of states (DOS) for bulk Fe-bcc and for Fe-2ML calculated using density functional theory. The Fermi level is $E_F$ = 0 eV.}
\label{F-dos-fe}
\end{figure}

To understand the reduction of $\sigma_{xx}$ in low dimensions, we studied the density of states (DOS) for bulk Fe-bcc and for Fe-2ML, for which the structures are the same as described above. As shown in Fig. \ref{F-dos-fe}, compared to the DOS of Fe-bcc, the DOS of Fe-2ML for spin up (majority spin) below the Fermi level ($E_F$ = 0 eV) becomes narrow and shifts down about 1.5 eV, and the DOS of Fe-2ML for spin down (minority spin) below the Fermi level becomes narrow and shifts up about 1.5 eV. The effect of the reduced dimensionality is a narrowing of the bands and an increase in the exchange splitting. This in turn decreases the diagonal part $\sigma_{xx}$ of the optical conductivity. 

\section{Discussion}
We briefly compare our results with those of the previous studies of the enhancement of the Kerr angle $\theta_{Kerr}$. 
It has been suggested that by considering alloys~\cite{Kerr-Sugimoto,DFT-Kerr-Osterloh} or multilayers~\cite{DFT-Kerr-Guo,DFT-Kerr-Ebert} of transition metals such as Fe or Co and heavy element such as Pt, $\theta_{Kerr}$ can be enhanced because heavy elements have large spin-orbit coupling. By contrast, we have found a way to enhance $\theta_{Kerr}$ and spin-orbit correlations {\it without using heavy elements} because avoiding the use of heavy elements is ideal for applications. Another mechanism for enhancing $\theta_{Kerr}$ is the plasma edge effect~\cite{DFT-Kerr-Feil-Haas}, which requires a dielectric constant satisfying the relation Re(${\epsilon_{xx}}$) = 1. Since $\theta_{Kerr}$ is proportional to $\epsilon_{xy}/(\epsilon_{xx}-1)$ as defined in Eq.(\ref{E-kerr}), the condition Re($\epsilon_{xx}$) = 1 induces a large $\theta_{Kerr}$ if Im($\epsilon_{xx}$) is small. Considering the relationship between optical conductivity and dielectric constant, $\sigma_{xx}=\frac{\omega}{4\pi i}(\epsilon_{xx}-1)$ given in Eq. (\ref{E-sigma}), the condition Re($\epsilon_{xx}$) = 1 and the condition Im($\sigma_{xx}$) = 0 are the same. Our study proposes a new method for enhancing $\theta_{Kerr}$ by considering Fe/insulator interfaces. Our study simultaneously considers both $\sigma_{xx}$ and $\sigma_{xy}$ on an equal footing, while the effect of $\sigma_{xy}$ has been ignored in studies that focused on the the plasma edge effect~\cite{DFT-Kerr-Feil-Haas}.       

\section{Conclusions}
In summary, using density functional theory calculations, we have found an enhanced MOKE in Fe/insulator multilayers. 
The Kerr effect is rather robust against the choice of insulator (vacuum, MgO, or AlF$_3$ in this study), but it requires the Fe layer to be atomically thin and to be dominated by interfacial Fe atoms. Our calculations suggest that the enhanced Kerr effect originates from the low-dimensional nature of the interfacial Fe atoms, which causes the following two effects: 
(i) The diagonal component $\sigma_{xx}$ of the optical conductivity decreases dramatically because the hopping integral for electrons between the Fe atoms is suppressed by the low dimensionality of the atomically thin Fe films. 
The off-diagonal component $\sigma_{xy}$ of the optical conductivity does not change at low photon energies, and it is enhanced at photon energies around 2 eV, where enhanced orbital magnetic moments and spin-orbit correlations are obtained for the interfacial Fe atoms. The Kerr angle $\theta_{Kerr}$ is proportional to the ratio $\sigma_{xy}/\sigma_{xx}$ by definition, and a large $\theta_{Kerr}$ develops at Fe/insulator interfaces with decreasing $\sigma_{xx}$ and large $\sigma_{xy}$. 
Our results show an efficient way to design novel devices, such as ultrathin ferromagnetic films~\cite{Okada} and ferromagnetic nanogranular films~\cite{Kobayashi-1,Kobayashi-2} by manipulating the substantial spin-orbit coupling effect at metal/insulator interfaces without using heavy elements. 
		
\section*{Acknowledgment}				
The authors acknowledge N. Kobayashi, K. Ikeda, and H. Masumoto for many valuable discussions about experiments on ferromagnetic nanogranular films. 	
		

\end{document}